\documentclass[aps,prd,twocolumn,nofootinbib,superscriptaddress]{revtex4-1} % revtex4.1 macros
\usepackage{amsmath,amssymb}
\usepackage{graphicx}
\usepackage{bigints}

\begin{document}

\title{Cosmological constraints from large-scale structure growth rate measurements}

%% Authors
\author{Anatoly Pavlov}
\email{pavlov@phys.ksu.edu}

\author{Omer Farooq}
\email{omer@phys.ksu.edu}

\author{Bharat Ratra}
\email{ratra@phys.ksu.edu}
\affiliation{Department of Physics, Kansas State University, 116 Cardwell Hall, Manhattan, Kansas 66506, USA}

\date{\today ~~~ KSUPT - 13/3}

\begin{abstract}

We compile a list of $14$ independent measurements of large-scale structure growth rate between redshifts $0.067 \leq z \leq 0.8$ and use this to place constraints on model parameters of constant and time-evolving general-relativistic dark energy cosmologies. With the assumption that gravity is well-modeled by general relativity, we discover that growth-rate data provide restrictive cosmological parameter constraints. In combination with type Ia supernova apparent magnitude versus redshift data and Hubble parameter measurements, the growth rate data are consistent with the standard spatially-flat $\Lambda$CDM model, as well as with mildly evolving dark energy density cosmological models.
\end{abstract}

\maketitle
\section{Introduction}
\label{Intro}

The discovery of the current acceleration of the cosmological expansion has raised the issue of whether this is due to a new form of matter -- dark energy -- or whether the general relativistic description of gravity needs to be modified. According to general relativity (GR), any form of energy affects space-time dynamics and so cosmological evolution. This fact allows for a very simple phenomenological explanation of the observed accelerated expansion, attributing it to a cosmological constant $\Lambda$, homogeneously distributed in space and constant in time. This $\Lambda$CDM model \citep{peebles84} is now accepted as the standard cosmological model. At the current epoch $\Lambda$ dominates the energy budget, with nonrelativistic cold dark matter (CDM) being the next largest contributor, followed by ordinary baryonic matter in third place. A widely discussed generalization of $\Lambda$CDM is the $\phi$CDM model in which $\Lambda$ is replaced by a time-varying dark energy density modeled by a self-interacting scalar field $\phi$ \citep{Peebles&Ratra1988}. For recent reviews see \cite{Tsujikawa2013}, \cite{Sola13}, \cite{Jimenez13}, and \cite{Burgess13}. An alternative explanation of the accelerated cosmological expansion is that GR is not the correct description of gravity on cosmological scales and must be modified so that on these large scales gravity has the property of making space expand with acceleration. For recent reviews of modified gravity see \cite{CapozzielloDeLaurentis2013} and \cite{SamiMyrzakulov2013}.

In this paper we assume that GR provides an adequate model for cosmological gravity and we test various models of dark energy (DE) as a possible explanation of the observed accelerated cosmological expansion. In particular, we consider three models of DE. The first one is the standard $\Lambda$CDM cosmology in which the energy density of DE does not evolve in time and its equation of state (EoS) is $p_\Lambda = -\rho_\Lambda$, where $p_\Lambda$ is the pressure and $\rho_\Lambda$ the energy density of DE. Space sections are not assumed to be flat in this case and the cosmological parameters that characterize the model are $\textbf{p} = (\Omega_{m0}, \Omega_{\Lambda})$ where $\Omega_{m0}$ is the current value of the nonrelativistic CDM and baryonic matter density parameter and $\Omega_{\Lambda}$ is that of $\Lambda$. The second model we consider is the simplest modification of $\Lambda$CDM cosmology in which the energy density of DE is time dependent and its EoS is parametrized as $p_{X} = w_{X}\rho_{X}$, where $w_X$ is constant and $< - 1/3$. The upper limit of $- 1/3$ is a consequence of the requirement that DE provide positive acceleration. This spatially-flat XCDM model is the simplest parametrization of dynamical DE, with parameters $\textbf{p} = (\Omega_{m0}, w_X)$. However, it is incomplete as it cannot describe density inhomogenities \citep[see, e.g.,][]{Podariu&Ratra2000}. The last model we study is the consistent quintessence model of DE in which DE is a scalar field. In particular we consider the much studied spatially-flat $\phi$CDM model \citep{Peebles&Ratra1988, Ratra&Peebles1988} whose equations of motion in units where $\hbar = c = 1$ are
\begin{eqnarray}
\label{phi_field_set}
\ddot\phi + 3\frac{\dot a}{a} \dot\phi -\frac{\kappa\alpha}{2}m_p^{\phantom{p}2} \phi^{-(\alpha+1)}&=&0,\\
\left(\frac{\dot a}{a}\right)^2&=&\frac{8\pi}{3m_p^{\phantom{p}2}}(\rho_m + \rho_\phi), \nonumber\\
\rho_\phi&=&\frac{m_p^{\phantom{p}2}}{32\pi}\left(\dot\phi^2 + \kappa m_p^{\phantom{p}2}\phi^{-\alpha}\right). \nonumber
\end{eqnarray}
Here an over-dot denotes a derivative with respect to time, $a$ is the scale factor, $\rho_m$ is the energy density of nonrelativistic (cold dark and baryonic) matter, $\rho_\phi$ is that of the dark energy scalar field $\phi$, $m_p = G^{-1/2}$ is the Planck mass where $G$ is the gravitational constant, and $\alpha > 0$ is a free parameter of the potential energy density of $\phi$ and determines $\kappa$ which is \citep[see][]{Peebles&Ratra1988, Ratra&Peebles1988}
\begin{eqnarray}
\label{kappa}
\kappa = \dfrac{8}{3}\left( \dfrac{\alpha + 4}{\alpha + 2} \right) \left[ \dfrac{2}{3}\alpha(\alpha + 2) \right]^{\alpha /2}.
\end{eqnarray}
In the limit $\alpha \longmapsto 0$ the $\phi$CDM model reproduces the spatially-flat $\Lambda$CDM cosmology while in the limit $\alpha \longmapsto \infty$ it reduces to the Einstein--De Sitter model with no DE but only CDM and baryonic matter. The value of $\alpha$ determines the rapidity of the time-evolution of the DE density, with a larger $\alpha$ corresponding to more rapidly decreasing DE density. The cosmological parameters of the $\phi$CDM model are $\textbf{p} = (\Omega_{m0}, \alpha)$.

Many different data sets have been used to derive constraints on the parameters of the three models we consider here.\footnote{See, e.g., \cite{SamushiaChenRatra2007}, \cite{ManiaRatra2012}, \cite{FarooqRatra2013a}, \cite{Piloyan2013}, \cite{Akarsu2014}, \cite{Li2013}, \cite{Perico2013}, and \cite{CardenasBernalBonila2013}; also see \cite{CrandallRatra2013}. For constraints on these and related models from near-future data see \cite{PavlovSamushia&Ratra2012}, \cite{Appleby&Linder2013}, \cite{Arabsalmani&Sahni2013}, and references therein.} In this paper we use growth-factor measurements to constrain cosmological parameters,\footnote{For related work with growth factor data, see \cite{Arkhipova&etal2014}.} under the assumption that GR is the correct model of gravity. Growth factor data have previously been used to test GR. Here we find that if we assume GR then growth-factor measurements provide tight constraints on cosmological parameters.

Our paper is organized as follows. In the next section we discuss the data and the analysis techniques that we use to derive cosmological parameter constraints. In Sec.\ \ref{results}, we present and discuss our results. Section\ \ref{concl} contains our conclusions.
%---------------------------------End of Introduction-------------------------------------------

\section{Data and analysis}
\label{data&analysis}

We use three different types of data to constrain cosmological parameters: the growth rate of large-scale structure (LSS) measurements; supernova type Ia (SNIa) distance-modulus measurements as a function of redshift; and Hubble parameter measurements.

\subsection{Growth rate of LSS}
\label{Growth}

%*******************************************Table 1************************************************************
\begin{table}[t]
\begin{center}
\begin{tabular}{cccc}
\hline\hline
~~$z$ & ~~$A(z)$ &~~~~~~~ $\sigma$ &~~ Reference\\
\tableline
0.067&~~	0.4230&~~~~~~~	0.0550&~~	1\\
0.150&~~	0.3900&~~~~~~~	0.0800&~~	2\\
0.170&~~	0.5100&~~~~~~~	0.0600&~~	3\\
0.220&~~	0.4200&~~~~~~~	0.0700&~~	4\\
0.250&~~	0.3512&~~~~~~~	0.0583&~~	5\\
0.350&~~	0.4400&~~~~~~~	0.0500&~~	3\\
0.370&~~	0.4602&~~~~~~~	0.0378&~~	5\\
0.410&~~	0.4500&~~~~~~~	0.0400&~~	4\\
0.550&~~	0.5000&~~~~~~~	0.0700&~~	6\\
0.570&~~	0.4150&~~~~~~~	0.0340&~~	7\\
0.600&~~	0.4300&~~~~~~~	0.0400&~~	4\\
0.770&~~	0.4900&~~~~~~~	0.1800&~~	8\\
0.780&~~	0.3800&~~~~~~~	0.0400&~~	4\\
0.800&~~	0.4700&~~~~~~~	0.0800&~~	9\\
\hline\hline
\end{tabular}
\end{center}
\caption{Growth parameter measurements and 1$\sigma$ uncertainties. Reference number shown in the last column:
1.\ \cite{Beutler2012}, 2.\ \cite{Hawkins2003}, 3.\ \cite{SongPercival2009}, 4.\ \cite{blake11}, 5.\ \cite{SamushiaPercivalRaccanelli2012}, 6.\ \cite{Ross2007}, 7.\ \cite{Reid2012}, 8.\ \cite{Guzzo2008}, 9.\ \cite{Torre2013}.}
\label{Tab_Az}
\end{table}
%*******************************************End of Table 1*****************************************************

In linear perturbation theory the nonrelativistic (cold dark and baryonic) matter density perturbation $\delta_m = \delta\rho_m/\rho_m$ obeys
\begin{eqnarray}
\label{growth_eq}
\ddot{\delta}_m + 2\dfrac{\dot{a}}{a}\dot{\delta}_m - \dfrac{4\pi}{m_p^{\phantom{p}2}}\rho_m\delta_m = 0,
\end{eqnarray}
where the scale factor $a$, with current value $a_0$, is related to redshift $z$ through $1 + z = a_0/a$. The analytic growing solution of (\ref{growth_eq}) is
\begin{eqnarray}
\label{Dz}
\delta_m(t) \propto D(z) = \dfrac{5\Omega_{m0}E(z)}{2}\int_{z}^{\infty}\dfrac{1 + z'}{E^{3}(z')}dz',
\end{eqnarray}
where $E(z) = H(z)/H_0$ and $H(z)$ is the Hubble parameter whose current value is the Hubble constant $H_0$. $D(z)$ is normalized such that $D(z = 0) = 1$. Note, that analytic solution (\ref{Dz}) is valid only for, in general spatially non-flat, $\Lambda$CDM cosmology. In cosmological models where dark energy density allowed to evolve in time, Eq.\ (\ref{growth_eq}) has to be solved numerically, which we do, in order to compute growth factor $D(z)$ for XCDM and $\phi$CDM cosmological models.

The observable we use in our analysis is constructed from the linear theory, redshift-dependent rms mass fluctuations in 8$h^{-1}$ Mpc spheres (where $h$ is $H_0$ in units of 100 km s$^{-1}$ Mpc $^{-1}$) $\sigma_8(z) = \sigma_8^0D(z)$, where $\sigma_8^0$ is the current value of $\sigma_8(z)$. We shall also need $f(z)$, the logarithmic derivative of the matter density perturbation $D(z)$ with respect to the scale factor $a$, $f(z) = d \ln D/d \ln a$. Using (\ref{Dz}) we find analytic expression for $f(z)$ we use to compute growth factor in $\Lambda$CDM cosmological model
\begin{eqnarray}
\label{Free_fz}
f(z) = \dfrac{\ddot{a}a}{\dot{a}^2} - 1 + \dfrac{5\Omega_{m0}}{2}\dfrac{(1 + z)^2}{E^{2}(z)D(z)}.
\end{eqnarray}
For XCDM and $\phi$CDM cosmological models we compute $f(z)$ numerically. The observable we use is the growth parameter $A_{\rm obs}(z) = f(z)\sigma_8(z)$ that also accounts for the Alcock-Paczynski effect in redshift-space distortions. The model prediction at redshift $z$ is $A_{\rm th}(z, \sigma_8^0, \textbf{p}) = f(z, \textbf{p})\sigma_8(z, \sigma_8^0, \textbf{p})$ where $\textbf{p}$ is the vector of cosmological parameters.

We use a $\chi^2$ analysis to derive constraints on cosmological parameters from growth factor data. $\chi^2$ depends on the cosmological parameters $\textbf{p}$ and $\sigma_8^0$,
\begin{eqnarray}
\label{chi^2_growth}
\chi^2_{G}(\sigma_8^0, \textbf{p}) = \sum_{i=1}^{N}\dfrac{[A_{\rm th}(z_i, \sigma_8^0, \textbf{p}) - A_{\rm obs}(z_i)]^2}{\sigma_{i}^2}.
\end{eqnarray}
Here $N$ is the number of data points and $\sigma_i$ is the 1$\sigma$ uncertainty on measurement $A_{\rm obs}(z_i)$ at redshift $z_i$, see Table \ref{Tab_Az}.\footnote{For the redshift $z = 0.57$ bin we use the value for model 2 from Table 1 of \cite{Reid2012} and an average of the upper and lower 1$\sigma$ uncertainties given for that model.} For our purposes, $\sigma_8^0$ is a nuisance parameter that we marginalize over. To do so we assume a Gaussian prior for $\sigma_8^0$ determined from cluster observations by \cite{Vikhlinin2009}, for spatially-flat $\Lambda$CDM, with mean $\overline{\sigma_8^0(\Omega_{m0})} = 0.813(\Omega_{m0}/0.25)^{-0.47}$ and 1$\sigma$ uncertainty $\sigma_{\overline{\sigma_8^0}}(\Omega_{m0}) = (\sigma_{\sigma_8^0}^2 + b^2)^{1/2}(\Omega_{m0}/0.25)^{-0.47}$, where the statistical uncertainty $\sigma_{\sigma_8^0} = 0.012$ and the systematic uncertainty $b = 0.02$ are added in quadrature. \cite{Vikhlinin2009} note that this relation is also adequate in the non-flat $\Lambda$CDM model and for alternative background cosmologies.\footnote{In this preliminary analysis we use this approximate, empirical expression for illustrative purposes. However, $\sigma_8^0$ does (weakly) depend on the full set of cosmological parameters $\textbf{p}$ in its own way for every cosmological model, so our analyses are approximate. Given that our results, described below, are encouraging, a more careful analysis that accounts for this effect is warranted and will be discussed elsewhere.} Then the posterior probability density function that depends only on the cosmological parameters $\textbf{p}$ is given by
\begin{widetext}

\begin{eqnarray}
\label{Lgrowth_Marg}
\mathcal{L}_{G}(\textbf{p}) = \frac{1}{\sigma_{\overline{\sigma_8^0}}(\Omega_{m0})\sqrt{2\pi}}&
\bigint \limits_{\!\!\!\!\!\!\!\!\! 0}^{\;\;\;\;\;\;\; \infty} \!{\rm exp}\left\lbrace  -\dfrac{\chi^2_{G}(\sigma_8^0,\textbf{p})}{2} - \dfrac{\left[ \sigma_8^0-\overline{\sigma_8^0(\Omega_{m0})} \right]^2}{2\sigma^{2}_{\overline{\sigma_8^0}}(\Omega_{m0})} \right\rbrace \, \mathrm{d}\sigma_8^0.
\end{eqnarray}

\end{widetext}
Finally, we compute the marginalized $\chi^2_{G}(\textbf{p}) = -2\ln(\mathcal{L}_{G}(\textbf{p}))$, and minimize this with respect to parameters $\textbf{p}$ to find the best-fit parameter values $\textbf{p}_0$. We also compute $1\sigma$, $2\sigma$, and $3\sigma$ cosmological parameter confidence contours bounded by $\chi^2_{G}(\textbf{p}) = \chi^2_{G}(\textbf{p}_0) + 2.3$, $\chi^2_{G}(\textbf{p}) = \chi^2_{G}(\textbf{p}_0) + 6.17$, and $\chi^2_{G}(\textbf{p}) = \chi^2_{G}(\textbf{p}_0) + 11.8$, respectively.

\subsection{SNIa distance modulus}
\label{SNIa}

The largest set of data we use are the 580 Type Ia supernova distance modulus $\mu_{\rm obs}(z)$ measurements from the \cite{suzuki12} Union 2.1 compilation (covering the redshift range of $0.015 \leq z \leq 1.414$). The predicted distance-modulus is
\begin{eqnarray}
\label{dist_modul}
\mu_{\rm th}(z) = 5\log_{10}[3000y(z)(1 + z)] + 25 - 5\log_{10}(h),
\end{eqnarray}
where $y(z)$ is the dimensionless coordinate distance
\begin{eqnarray}
\label{yz}
y(z) = \frac{1}{\sqrt{-\Omega_k}} ~\mathrm{sin}\left(\sqrt{-\Omega_k}
\int \limits_{0}^{z}{\frac{dz'}{E(z')}}\right),
\end{eqnarray}
and $\Omega_k$ is the spatial curvature density parameter. Since the SNIa distance modulus measurements $\mu_{\rm obs}$ are correlated we use $\chi^2$ defined through the inverse covariance matrix $\chi_{SN}^{2}(h,\textbf{p})=\Delta\boldsymbol\mu^T~ C^{-1}~\Delta\boldsymbol{\mu}$. Here the vector of differences $\Delta{\mu_i}= \mu_{\rm th}(z_i, H_0, \textbf{p}) -\mu_{\rm obs}(z_i)$, and $C^{-1}$ is the inverse of the 580 by 580 Union 2.1 compilation covariance matrix.

\subsection{Hubble parameter}
\label{Hz}

We use 20 Hubble parameter measurements $H_{\rm obs}(z)$ and 1$\sigma$ uncertainties covering redshift range $0.09 \leq z \leq 2.3$ \citep{simon05, Stern2010, moresco12, busca12}, as listed in Table 1 of \cite{FarooqRatra2013b}. We only include independent measurements of the Hubble parameter, i.e., we exclude $H_{\rm obs}(z)$ points that are possibly correlated with growth factor measurements in Table\ \ref{Tab_Az} above.

Theoretical expressions for the Hubble parameter follow directly from the Friedmann equation in each model. In the case of the $\Lambda$CDM model,
\begin{eqnarray}
\label{Hz_th_LCDM}
&H&_{\rm th}^2(z, \textbf{p}) =\nonumber\\
&H&_0^2\left[ \Omega_{m0}(1 + z)^3 + (1 - \Omega_{m0} - \Omega_{\Lambda})(1 + z)^2 + \Omega_{\Lambda} \right],
\end{eqnarray}
while for the spatially-flat XCDM parameterization, 
\begin{eqnarray}
\label{Hz_th_XCDM}
&H&_{\rm th}^2(z, \textbf{p}) =\nonumber\\
&H&_0^2\left[ \Omega_{m0}(1 + z)^3 + (1 - \Omega_{m0})(1 + z)^{3(1 + w_X)} \right],
\end{eqnarray}
and for the $\phi$CDM model,
\begin{eqnarray}
\label{Hz_th_phiCDM}
H_{\rm th}^2(z, \textbf{p}) &=& H_0^2\Omega_{m0}(1 + z)^3\nonumber\\
&+& \frac{1}{12}\left(\dot{\phi}^{2} + \kappa m_p^{\phantom{p}2}\phi ^{-\alpha} \right).
\end{eqnarray}

We use the same technique to constrain cosmological parameters from $H(z)$ measurements as we used in Sec.\ \ref{Growth} for the growth factor data analysis. First, we define $\chi^2_{H}(H_0, \textbf{p})$ in accordance with Eq.\ (\ref{chi^2_growth}) where instead of the growth factor $A(z)$ we insert the Hubble parameter $H(z)$.

\subsection{Computation of joint $\chi^2(\textbf{p})$}
\label{joint_chi2}

We perform two joint analyses, one for the combination of SNIa and $H(z)$ data, the other for all three data sets. For the SNIa+$H(z)$ analysis we multiply likelihood functions from the SNIa data and the $H(z)$ data and then marginalize this over the nuisance parameter $H_0$ with a Gaussian prior with mean value $\overline{H_0} = 68.0$ km s$^{-1}$ Mpc $^{-1}$ and 1$\sigma$ uncertainty $\sigma_{H_0} = 2.8$ km s$^{-1}$ Mpc $^{-1}$ (\cite{ChenRatra2011a}, also see \cite{Gott2001}, \cite{ChenGott&Ratra2003}, \cite{Calabrese2013}) to finally determine the joint
$\chi^2_{\mathrm{SNIa+} H}(\textbf{p})$ function, which depends only on cosmological parameters $\textbf{p}$. This is then used to find the best-fit values of $\textbf{p}_0$ and corresponding cosmological parameter constraints. The second joint analysis, of the SNIa+$H(z)$ data with the growth factor data, is based on adding their $\chi^2$-functions, $\chi^2_{\rm Jnt}(\textbf{p}) = \chi^2_{\mathrm{SNIa+} H}(\textbf{p}) + \chi^2_{G}(\textbf{p})$.
%---------------------------------End of Data and Analysis--------------------------------------

\section{Results and discussion}
\label{results}

\begin{figure*}
\centering
  \includegraphics[width=1.0\textwidth]{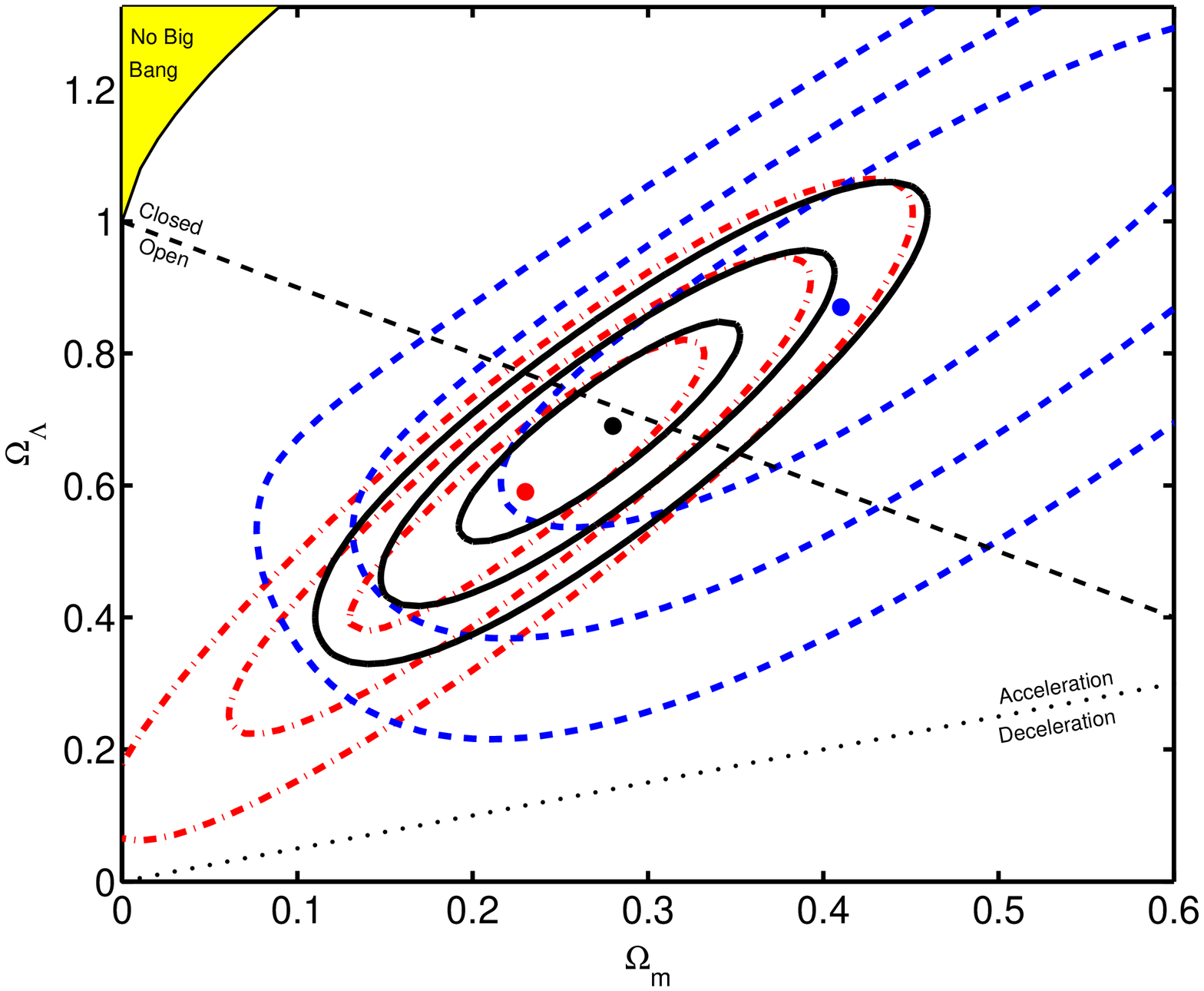}
\caption{
1, 2, and 3$\sigma$ constraint contours for the $\Lambda$CDM model from: growth factor measurements [blue dashed lines with blue filled circle at best-fit ($\Omega_m$, $\Omega_\Lambda$) = (0.41, 0.87) with $\chi_{\rm min}^{2}/\rm dof$ = 7.65/12]; SNIa+$H(z)$ apparent magnitude data [red dot-dashed lines with red filled circle at best-fit ($\Omega_m$, $\Omega_\Lambda$) = (0.23, 0.59) with $\chi_{\rm min}^{2}/\rm dof$ = 562/598]; and a combination of all data sets [black solid lines and black filled circle at best-fit ($\Omega_m$, $\Omega_\Lambda$) = (0.28, 0.69) with $\chi_{\rm min}^{2}/\rm dof$ = 571/612]. The dashed straight line corresponds to spatially-flat models, the dotted line demarcates zero acceleration models, and the area in the upper left-hand corner is the region for which there is no big bang.
}
\label{fig:LCDM}
\end{figure*}

\begin{figure*}
\centering
  \includegraphics[width=1.0\textwidth]{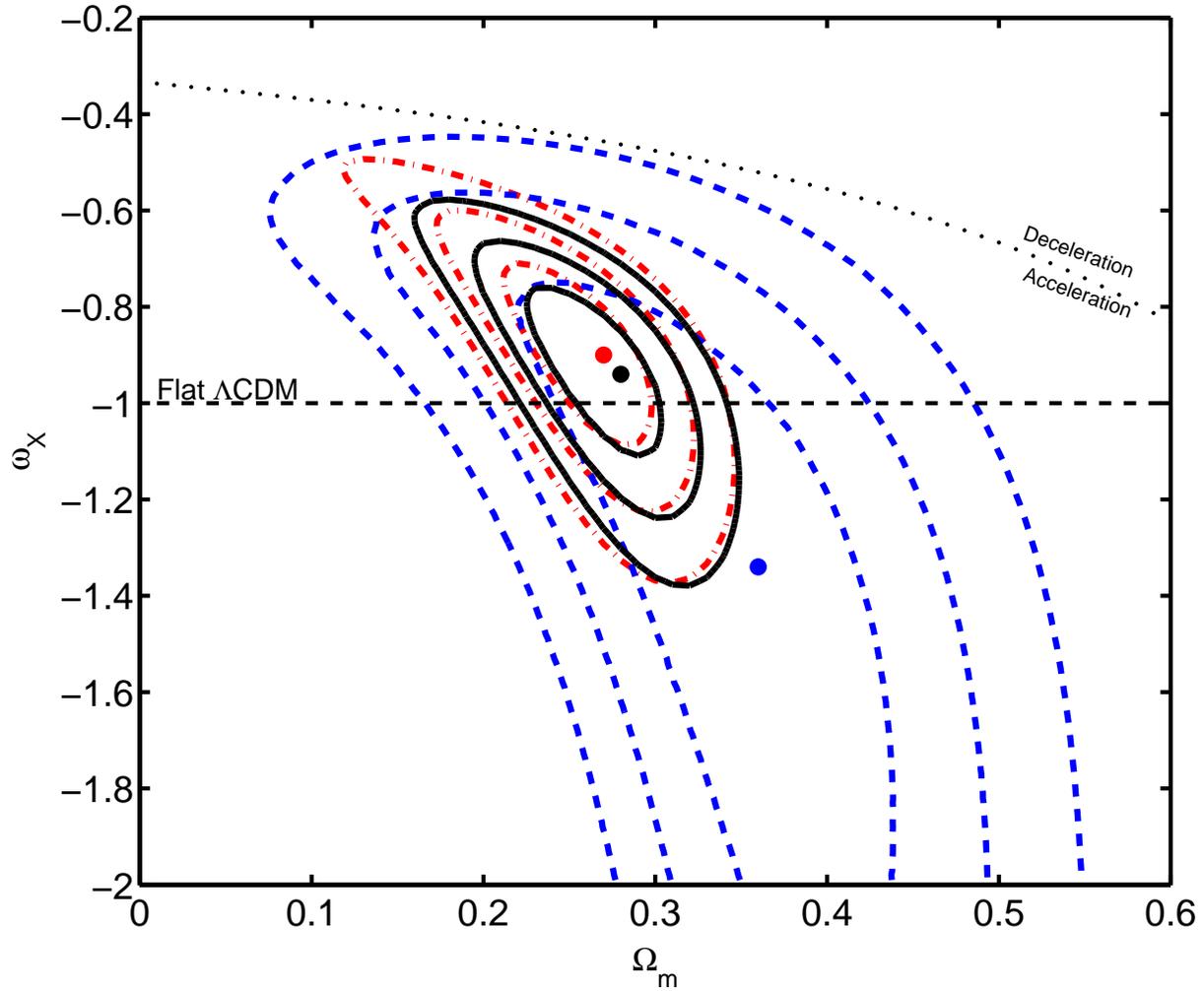}
\caption{
1, 2, and 3$\sigma$ constraint contours for the XCDM model from: growth factor measurements [blue dashed lines with blue filled circle at best-fit ($\Omega_m$, $w_X$) = (0.36, -1.34) with $\chi_{\rm min}^{2}/\rm dof$ = 7.70/12]; SNIa+$H(z)$ apparent magnitude data [red dot-dashed lines with red filled circle at best-fit ($\Omega_m$, $w_X$) = (0.27, -0.90) with $\chi_{\rm min}^{2}/\rm dof$ = 562/598]; and a combination of all data sets [black solid lines and black filled circle at best-fit ($\Omega_m$, $w_X$) = (0.28, -0.94) with $\chi_{\rm min}^{2}/\rm dof$ = 571/612]. The dashed straight line corresponds to spatially-flat $\Lambda$CDM models and the dotted curved line demarcates zero acceleration models.
}
\label{fig:XCDM}
\end{figure*}

\begin{figure*}
\centering
  \includegraphics[width=1.0\textwidth]{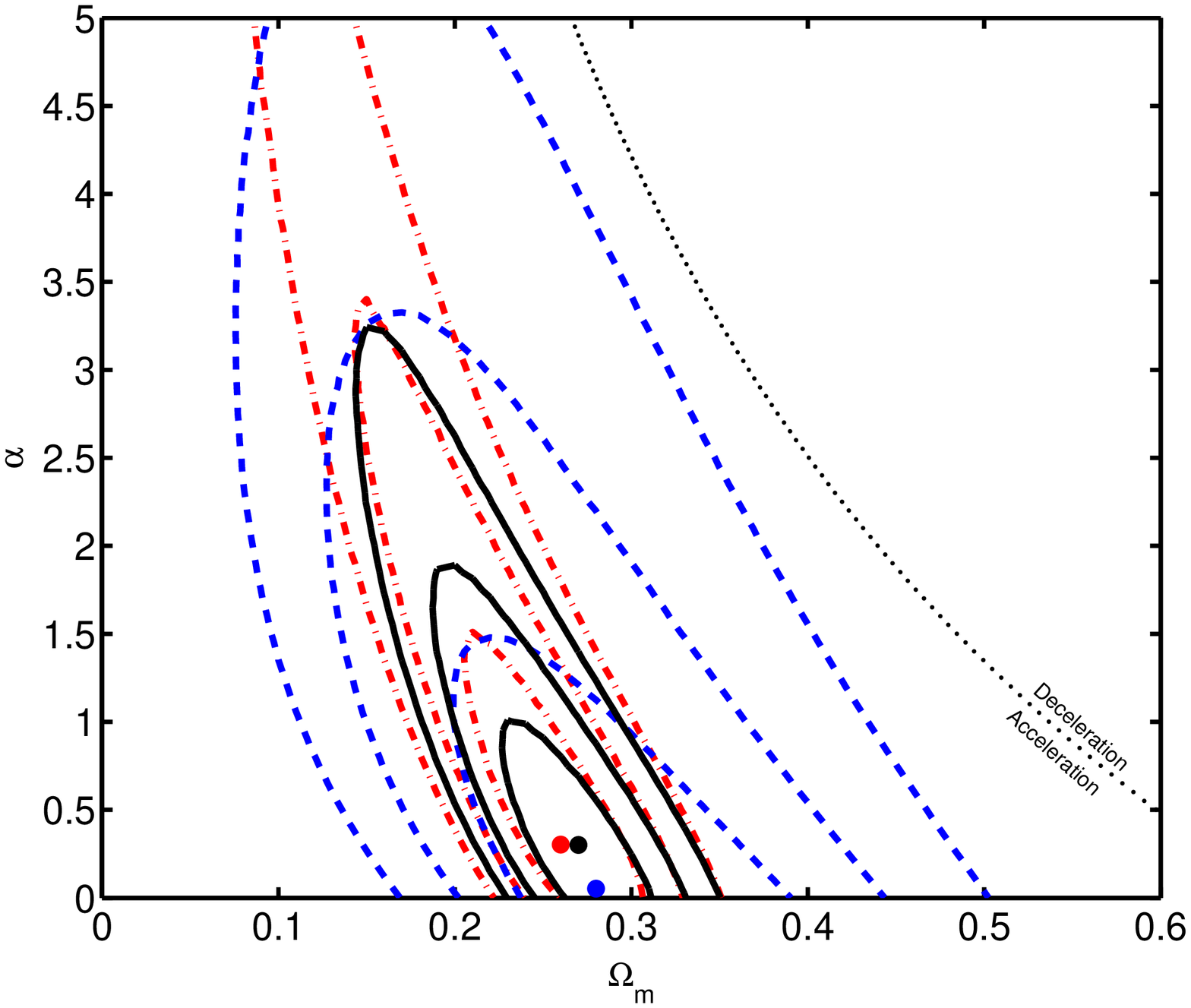}
\caption{
1, 2, and 3$\sigma$ constraint contours for the $\phi$CDM model from: growth factor measurements [blue dashed lines with blue filled circle at best-fit ($\Omega_m$, $\alpha$) = (0.28, 0.052) with $\chi_{\rm min}^{2}/\rm dof$ = 8.62/12]; SNIa+$H(z)$ apparent magnitude data [red dot-dashed lines with red filled circle at best-fit ($\Omega_m$, $\alpha$) = (0.26, 0.302) with $\chi_{\rm min}^{2}/\rm dof$ = 562/598]; and a combination of all data sets [black solid lines and black filled circle at best-fit ($\Omega_m$, $\alpha$) = (0.27, 0.300) with $\chi_{\rm min}^{2}/\rm dof$ = 570/612]. The dotted curved line demarcates zero acceleration models and the horizontal $\alpha = 0$ axis corresponds to spatially-flat $\Lambda$CDM models.
}
\label{fig:phiCDM}
\end{figure*}

We derived cosmological parameter constraints from combination of SNIa+$H(z)$ data sets as well as from a joint analysis of all three data sets.\footnote{We have not used all the $H(z)$ measurements in this paper, excluding points from Table 1 of \cite{FarooqRatra2013b} that are possibly correlated with some of the growth rate data we use in this paper.} Our results, presented in the form of isocontours in cosmological parameter space, are shown in Figs.\ \ref{fig:LCDM}, \ref{fig:XCDM}, and \ref{fig:phiCDM} for the $\Lambda$CDM, XCDM and $\phi$CDM models, respectively.

In the $\Lambda$CDM model the growth factor data favor higher best-fit value of a negative spatial curvature parameter $\Omega_{k0} = 1 - \Omega_{m0} - \Omega_\Lambda$ (which corresponds to a closed, spherical spatial geometry) along with a higher best-fit value of $\Omega_{m0}$ compared to what other cosmological tests favor, such as SNIa, Hubble parameter measurements, BAO and CMB (see for example \cite{SamushiaChenRatra2007}-\cite{CrandallRatra2013} and references therein). In the case of the XCDM parameterization the growth factor data favor a steeper time dependence of dark energy density and also a higher value of ordinary matter energy density parameter (i.e. equation of state parameter $w_X$ has a lower best-fit value and $\Omega_{m0}$ has a higher best-fit value) in comparison with constraints derived from the above-mentioned data sets. However, for the $\phi$CDM model one observes consistent results for the best-fit values of cosmological parameters ($\Omega_{m0}$, $\alpha$) with those previously obtained using the data sets mentioned above.

Also, our results for the $\Lambda$CDM model differ from constraints obtained for this model from other analyses of growth factor data (see \cite{Beutler2012}-\cite{Reid2012}). We suspect that the reason for this and the reason that the constraining power of growth rate data has not previously been recognized is because these data have almost always been used to constrain cosmological parameters in the context of modified gravity models. These modified gravity models have more free parameters than the models we have considered here, because we have assumed that general relativity provides an adequate description of gravitation on cosmological scales.

The other striking feature of the growth rate data constraints is that for all three models they align well with those of the SNIa+$H(z)$ joint constraints.
%-----------------------------------------End of Results-----------------------------------------------

\section{Conclusion}
\label{concl}

We have used three general-relativistic DE cosmological models to analyze the largest collection of growth factor measurements to date. We have discovered that growth factor data constraints on cosmological parameters are quite restrictive, roughly close to those from joint SNIa+$H(z)$ and baryon acoustic oscillations (BAO) peak length-scale measurements, and less restrictive than those from cosmic microwave background (CMB) anisotropy observations.

These growth factor results must be viewed as tentative, given that this is an area of research that is still under active development. It is important to continue to study possible sources of systematic uncertainty -- and given the differences we have found between growth rate data constraints and these from SNIa and $H(z)$ measurements, it is not unreasonable to suspect that there might be an as yet hidden source of systematic uncertainty.

It is, however, clear that growth factor measurements will soon be able to provide cosmological constraints as restrictive and as reliable as those from CMB anisotropy, BAO, $H(z)$, and SNIa measurements.
%---------------------------------End of Conclusion--------------------------------------------
\acknowledgements

We thank Lado Samushia for very helpful suggestions and discussions about growth rate measurements that helped us choose the data for our analyses. We thank Mikhail Makouski for useful discussions and valuable advice about numerical techniques. This work was supported in part by DOE Grant No.\ DEFG03-99EP41093 and NSF Grant No.\  AST-1109275.
%---------------------------------End of Acknowledgments--------------------------------------

%Bibliography----------------------------------------------------------------------------------

%********************************************Figures***********************************************************
%Fig.1

%Fig.2

%Fig.3

\end{document}